\documentstyle[12pt,axodraw]{article}
\newenvironment{comment}[1]{}{}
\def\beq{\begin{equation}}
\def\eeq{\end{equation}}
\def\bea{\begin{eqnarray}}
\def\eea{\end{eqnarray}}
\def\barr{\begin{array}}
\def\earr{\end{array}}
\newcommand{\sm}{standard model}
\newcommand{\cm}{center of mass}
\newcommand{\xs}{cross section}
\newcommand{\lc}{linear collider}

\begin{document}

\title{\Large\bf\boldmath
MANIFESTATIONS OF STRONG ELECTROWEAK SYMMETRY BREAKING IN $e^-e^-$ SCATTERING
\unboldmath}
\author{{\large FRANK CUYPERS}\\\\
{\tt cuypers@mppmu.mpg.de}\\
{\em Max-Planck-Institut f\"ur Physik, Werner-Heisenberg-Institut,}\\
{\em D-80805 M\"unchen, Germany}
}
\date{}
\maketitle
\pagestyle{empty}
\thispagestyle{empty}
\begin{abstract}
\noindent\footnotesize
We analyze the incidence
on polarized $e^-e^-$ scattering
of the trilinear and quartic anomalous gauge couplings
which arise in the realm of a non-linear realization
of electroweak symmetry breaking.
\end{abstract}
\vfill

\section{Introduction}

The tree-level gauge couplings in the \sm\
are dictated by the gauge principle.
Deviations from this expectation
can arise from several sources,
including
quantum corrections, new particles or compositeness.
Since the \sm\ quantum corrections are expected to be small~\cite{couture},
such deviations would provide stringent evidence for new physics.
Unfortunately,
the present experimental bounds
still lack by far the accuracy necessary to detect any effect.
Dramatic improvements,
though,
are expected from LHC~\cite{bdv}
or the next generation of \lc s operated
in the $e^+ e^-$~\cite{b},
$e^-\gamma$ and $\gamma\gamma$~\cite{photon}
or $e^-e^-$~\cite{cc,ck} modes.

We consider here a heavy Higgs scenario,
where the electroweak symmetry breaking sector is strongly coupled~\cite{lw}.
Chiral perturbation theory enables us
to write an effective lagrangian
in which the symmetry breaking pattern is fixed
by a custodial $SU(2)_c$ global symmetry.
The heavy
(or even non-existent)
Higgs
is then seen through an infinite tower of operators,
of which only those of lowest order
in a momentum expansion
contribute to the low energy effective theory.
The coefficients of these operators
parametrize the unknown physics,
which by assumption lies outside the reach of direct investigations.
In the unitary gauge,
the \sm\ lagrangian is then supplemented by
the following effective lagrangians~\cite{bdv,b}:

\clearpage
\bea
L_T &=~
-i \displaystyle{e^3 \over 32\pi^2 s_w^2}&
\Biggl[
    L_{9R}
    \left(
      W^\dagger_\mu W_\nu A^{\mu\nu}
    - {s_w \over c_w} W^\dagger_\mu W_\nu Z^{\mu\nu}
    \right)
\label{lt}\\&&\nonumber
  +
    L_{9L}
    \left(
      W^\dagger_\mu W_\nu A^{\mu\nu}
    + {c_w \over s_w} W^\dagger_\mu W_\nu Z^{\mu\nu}
    + {1 \over s_wc_w} \left(W^\dagger_{\mu\nu}W^\mu
      - W_{\mu\nu}W^{\dagger\mu}\right) Z^{\nu}
    \right)
\Biggr]
\\\nonumber\\
L_Q &= ~
\displaystyle{e^4 \over 32\pi^2 s_w^4}&
\Biggl[
    L_1
    \left(
      2(W^\dagger \!\cdot\! W)^2 + {2 \over c_w^2} (W^\dagger \!\cdot\! W) Z
\!\cdot\! Z + {1 \over 2c_w^2} (Z \!\cdot\! Z)^2
    \right)
\label{lq}\\&&\nonumber
  + L_2
    \left(
      (W^\dagger \!\cdot\! W)^2 + |W \!\cdot\! W|^2 + {2 \over c_w^2}
(W^\dagger \!\cdot\! W) Z \!\cdot\! Z + {1 \over 2c_w^2} (Z \!\cdot\! Z)^2
    \right)
\Biggr]~,
\eea
where
$W_{\mu\nu} = \partial_\mu W_\nu - \partial_\nu W_\mu$
and
$V_{\mu\nu} = \partial_\mu V_\nu - \partial_\nu V_\mu$
($V = \gamma,Z$).

The lagrangian $L_T$
describes anomalous trilinear interactions among the gauge bosons,
whereas $L_Q$
modifies the quartic couplings.
The latter is also responsible for a new type of $ZZZZ$ interaction,
which is absent in the \sm.

It should be remembered that the coefficients
$L_{9L}$, $L_{9R}$, $L_1$ and $L_2$
of the anomalous operators
are no fundamental constants,
but merely energy-dependent form factors.
Moreover,
the operators responsible for $L_T$
typically originates form loops,
whereas $L_Q$ can be generated at tree-level~\cite{aew}.
Therefore,
the natural expectation
is that the coefficients $L_{9L}$ and $L_{9R}$
are small compared to $L_1$ and $L_2$.
Note also that the operators from which $L_T$ is derived
induce quartic terms as well.
These, however, should be small compared to the ``genuine''
quartic anomalies
of $L_Q$,
because of the argument given above.
We therefore ignore them from now on.

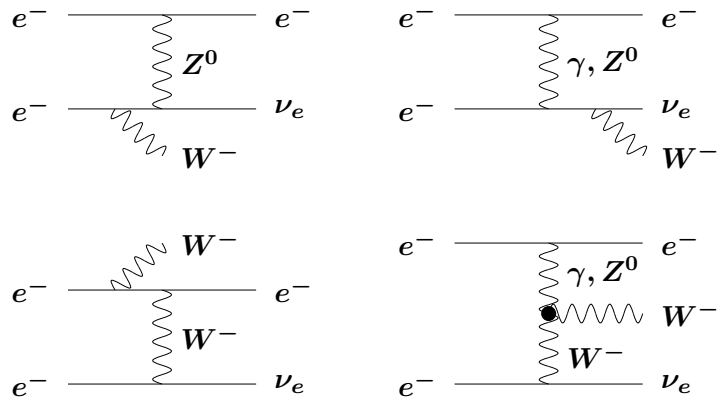
\begin{figure}[htb]
\unitlength.25mm
\SetScale{.709}
\boldmath
\begin{picture}(200,100)(-180,0)
\Line(0,25)(100,25)
\Line(0,75)(100,75)
\Photon(50,25)(50,75){5}{5}
\Photon(25,25)(50,0){-5}{4}
\Text(-10,25)[r]{\small$e^-$}
\Text(-10,75)[r]{\small$e^-$}
\Text(110,25)[l]{\small$\nu_e$}
\Text(110,75)[l]{\small$e^-$}
\Text(60,50)[l]{\small$Z^0$}
\Text(60,0)[l]{\small$W^-$}
\end{picture}
\begin{picture}(200,80)(-180,0)
\Line(0,25)(100,25)
\Line(0,75)(100,75)
\Photon(50,25)(50,75){5}{5}
\Photon(75,25)(100,0){-5}{4}
\Text(-10,25)[r]{\small$e^-$}
\Text(-10,75)[r]{\small$e^-$}
\Text(110,25)[l]{\small$\nu_e$}
\Text(110,75)[l]{\small$e^-$}
\Text(60,50)[l]{\small$\gamma,Z^0$}
\Text(110,0)[l]{\small$W^-$}
\end{picture}
\\
\begin{picture}(200,120)(-180,0)
\Line(0,0)(100,0)
\Line(0,50)(100,50)
\Photon(50,0)(50,50){5}{5}
\Photon(25,50)(50,75){5}{4}
\Text(-10,0)[r]{\small$e^-$}
\Text(-10,50)[r]{\small$e^-$}
\Text(110,0)[l]{\small$\nu_e$}
\Text(110,50)[l]{\small$e^-$}
\Text(60,75)[l]{\small$W^-$}
\Text(60,25)[l]{\small$W^-$}
\end{picture}
\begin{picture}(00,120)(-180,0)
\Line(0,0)(100,0)
\Line(0,75)(100,75)
\Photon(50,0)(50,75){5}{8}
\Photon(50,37.5)(100,37.5){5}{5}
\Vertex(50,37.5){4}
\Text(-10,0)[r]{\small$e^-$}
\Text(-10,75)[r]{\small$e^-$}
\Text(110,0)[l]{\small$\nu_e$}
\Text(110,75)[l]{\small$e^-$}
\Text(60,15)[l]{\small$W^-$}
\Text(60,60)[l]{\small$\gamma,Z^0$}
\Text(110,37.5)[l]{\small$W^-$}
\end{picture}
\unboldmath
\caption{\footnotesize
  Typical lowest order Feynman diagrams
  participating to $W^-$ production in $e^-e^-$ scattering.
}
\label{feyn}
\end{figure}

We analyze here how the addition of the pieces (\ref{lt},\ref{lq})
to the \sm\ lagrangian
modifies the latter's predictions in $e^-e^-$ scattering.
For this we examine the $W^-$ and $W^-Z^0$ production reactions,
which respectively probe the
trilinear and quartic gauge couplings.

\section{Trilinear Couplings}

In $e^-e^-$ scattering the trilinear anomalous couplings
$L_{9L}$ and $L_{9R}$
appear to lowest order in the reaction
\beq
e^-e^- \to e^-\nu_eW^-
\label{eq1}~.
\eeq
The Feynman diagrams contributing to this process
are depicted in Fig.~\ref{feyn}.
It is the last diagram,
whose trilinear vertex is emphasized,
which provides the signals we intend to test.
If both electron beams are right-polarized
the process does not take place,
because at least one of the fermion lines
is connected to a $W$ boson.
For the $LR$ combination of initial helicities,
the third diagram
in which a $W$ is exchanged between the two fermion lines,
does not contribute.
For the $LL$ configuration,
all diagrams contribute.

\begin{figure}[htb]
\boldmath
\input{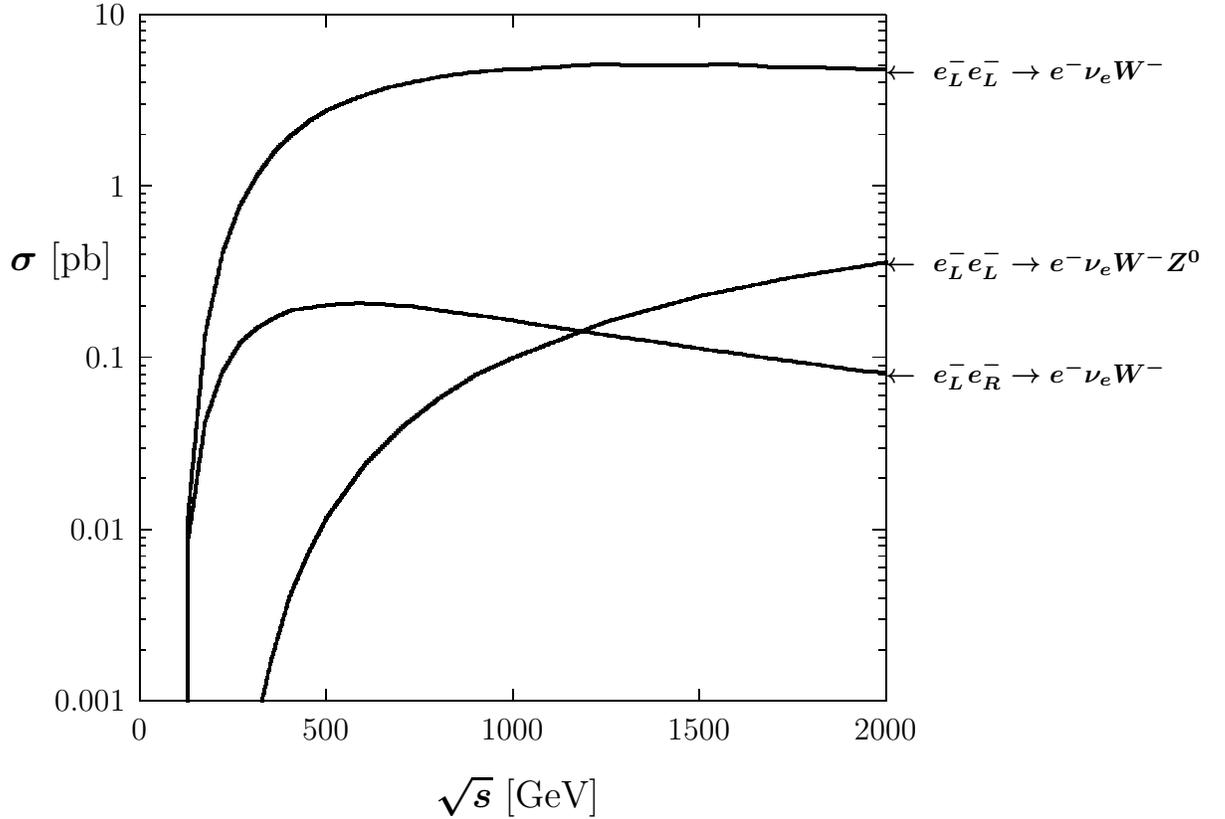}
\unboldmath
\caption{\footnotesize
  Energy dependence of the polarized $W^-$ production \xs s
  considered in the text.
}
\label{eny}
\end{figure}

To make sure
these and only these $e^-\nu_eW^-$ events are seen,
we impose the following cuts on the final state:
\renewcommand{\arraystretch}{1.5}
\beq
\left|
\begin{array}{l}
\theta_e > 10^o
\\
E_e > 10 \mbox{ GeV}
\\
p_\perp^{\rm vis} > 10 \mbox{ GeV}
\\
\mbox{efficiency}={2\over3} \quad\Leftarrow\quad \mbox{only hadronic decays of
the } W~.
\end{array}
\right.
\label{cuts}
\eeq
\renewcommand{\arraystretch}{1}

When combined with the measurement of the jets invariant mass,
the restriction to only hadronic decays of the $W$
insures the absence of non-resonant backgrounds.
Similarly,
the requirement of an imbalance in the transverse momentum
guarantees that no photoproduction events
with only one electron lost along the beam pipe
are included.
The expected \sm\ \xs s~\cite{cc}
are displayed as functions of the \cm\ energy in Fig.~\ref{eny}.
In all calculations we ignore the electron mass,
as it is justified at the energies considered
with the cuts (\ref{cuts}).

\begin{figure}[htb]
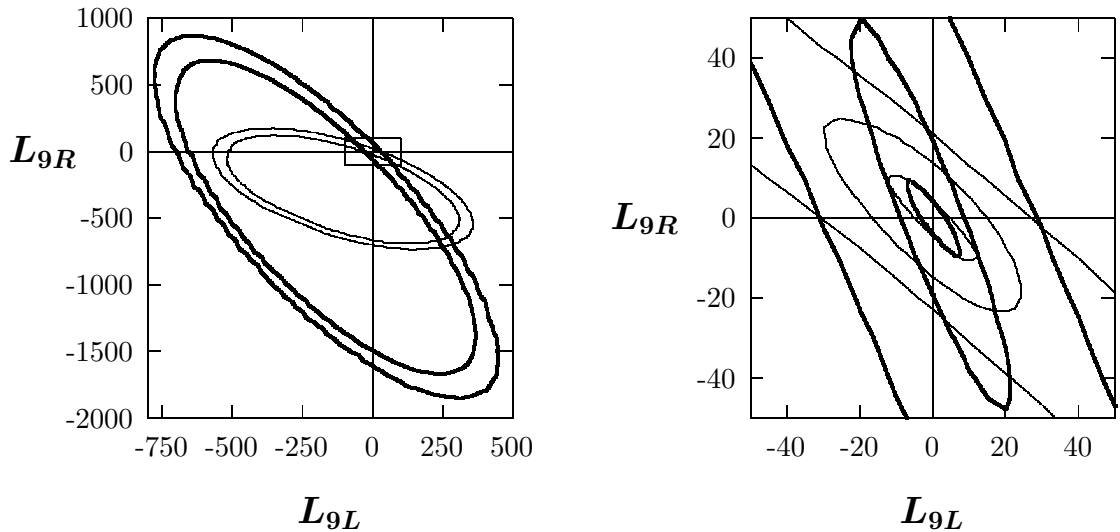

\boldmath
\centerline{\small
\input{glo.500.tex}
\input{loc.500.tex}
}
\unboldmath
\caption{\footnotesize
  Contours of detectability
  at 95\%\ confidence
  for the trilinear anomalous couplings
  in the polarized $e^-e^- \to e^-\nu_eW^-$ reactions,
  with the \cm\ energy $\protect\sqrt{s}=500$ GeV
  and the luminosity ${\cal L}=20$ fb$^{-1}$.
  The thin curves are for the $LR$ combination
  of initial electron polarizations,
  whereas the thick curves are for the $LL$ combination.
  The outer to inner contours
  correspond to the three different analysis
  described in the text.
  The plot on the right is a zoom
  into the boxed region of the left plot.
}
\label{fg500}
\end{figure}

To determine the discovery potential of the reaction (\ref{eq1})
we have used least squares estimators
as in Ref.~\cite{cc}.
Assuming there is no anomalous coupling,
we explore this way the region around
$L_{9L} = L_{9R} = 0$
for the finite values of these parameters
which can be excluded
with 95\%\ confidence
($\chi^2\ge6$).
We have performed this analysis with three procedures
of increasing resolving power:
\clearpage\noindent
\begin{enumerate}
\item
  Using only the information from the total \xs s.
  As we shall see,
  this is totally inadequate.
\item
  Adding information from the differential \xs s,
  by subdividing the emerging electron's polar angle range
  into 20 bins for the $LL$ case and
  10 bins for the $LR$ case.
  This way each bin is guaranteed to contain a sufficiently
  large number of nearly gaussian distributed events.
\begin{comment}{
  \footnote{
    This is in order to guarantee
    that even in the $LR$ case
    each bin contains a sufficiently
    large number of nearly gaussian distributed events.
  }
}\end{comment}
\item
  Computing the Cramer-Rao limit of this reaction~\cite{cr}
  \beq
  \chi_\infty^2 = {\cal L} \int d\Omega ~
  \displaystyle{
    \left( \Delta {d\sigma/d\Omega} \right)^2
    \over
    {d\sigma/d\Omega}
    }~,
  \eeq
  where $d\Omega$ is the element of phase space.
  The resulting bounds on the parameters
  are the best one may hope to ever achieve
  with a given luminosity $\cal L$
  in a perfect experiment.
  They may therefore serve as a benchmark of the reaction.
\end{enumerate}

The number of events
is calculated assuming a typical integrated design luminosity
scaling like
${\cal L} = 80s\mbox{ fb}^{-1}/\mbox{TeV}$.
Since the polar angle of an electron should be measurable
to an accuracy better than 10 mrad,
the main systematic errors originate
from the luminosity measurement
and from detector inefficiencies.
None of these should exceed 1\%,
and we conservatively assume an overall systematic uncertainty of 2\%.

\vspace*{-8mm}
\begin{figure}[htb]
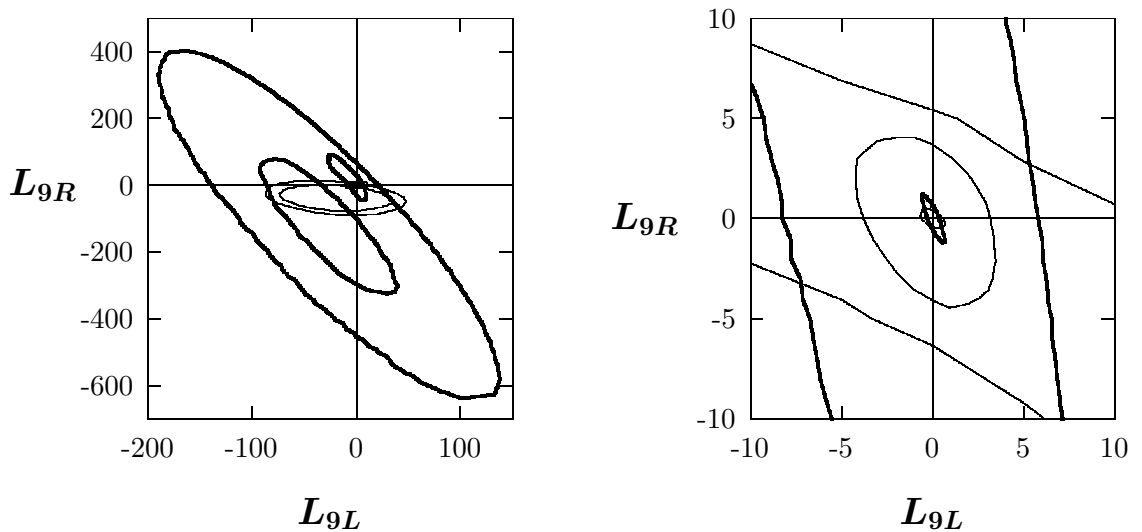

\boldmath
\centerline{\small
\input{glo.2000.tex}
\input{loc.2000.tex}
}
\unboldmath
\caption{\footnotesize
  Same as Fig.~\protect\ref{fg500},
  with $\protect\sqrt{s}=2$ TeV
  and ${\cal L}=320$ fb$^{-1}$.
}
\label{fg2000}
\end{figure}

In the second analysis,
instead of the electron's polar angle
one could as well have chosen another variable,
such as the polar angles of the $W^-$ or the $\nu_e$,
or the energy or transverse momentum of any combination of the particles.
As it turns out,
the polar angle of the $W^-$ is actually
a more sensitive variable,
but it cannot be measured as precisely as the angle of the electron.

In Figs~\ref{fg500} and \ref{fg2000}
we have plotted the 95\%\ observability contours
of $L_{9L}$ and $L_{9R}$
which can be obtained in 500 GeV and 2 TeV \cm\ collisions
with each of the three procedures.
The information from the total \xs\ is clearly not sufficient,
since it cannot resolve an ambiguity along a closed curve
in these two parameters.
Taking into account the electron's angular correlations,
though,
improves the resolution dramatically
and totally lifts the ambiguity\footnote{
  Note that a two-fold ambiguity always subsists
  when $L_{9L}$ and $L_{9R}$ are probed
  with the $e^+e^- \to W^+W^-$ reaction.
}.
There may,
however,
be much more to be gained
by using improved observables
and multivariate correlations
or a maximum likelihood fitting,
since the theoretical Cramer-Rao limit of this reaction
is still doing better by several factors of one.

It is worth noting
that although the \sm\ \xs s
for the $LR$ combination of beam polarizations
are much less than those of the $LL$ combination,
this experiment can provide better bounds on the parameters.
Indeed,
the large \xs s of the $LL$ mode
are also due to the many more background diagrams.
The resolving power of a reaction
is not necessarily related to the \sm\ rates.

\section{Quartic Couplings}

\begin{figure}[htb]
\begin{center}

\unitlength.25mm
\SetScale{.709}

\fbox{
\begin{picture}(95,95)
\Line(20,20)(80,20)
\Line(20,60)(80,60)
\Photon(50,20)(50,60){2}{8}
\Photon(50,40)(80,40){2}{4}
\Photon(35,60)(55,80){2}{4}
\end{picture}
}
\fbox{
\begin{picture}(95,95)
\Line(20,20)(80,20)
\Line(20,60)(80,60)
\Photon(50,20)(50,60){2}{8}
\Photon(50,40)(80,40){2}{4}
\Photon(65,60)(85,80){2}{4}
\end{picture}
}
\fbox{
\begin{picture}(95,95)
\Line(20,20)(80,20)
\Line(20,80)(80,80)
\Photon(50,20)(50,80){2}{12}
\Photon(50,50)(60,50){2}{2}
\Photon(60,50)(80,40){2}{4}
\Photon(60,50)(80,60){2}{4}
\end{picture}
}
\fbox{
\begin{picture}(95,95)
\Line(20,20)(80,20)
\Line(20,80)(80,80)
\Photon(50,20)(50,80){2}{12}
\DashLine(50,50)(60,50){2}
\Photon(60,50)(80,40){2}{4}
\Photon(60,50)(80,60){2}{4}
\end{picture}
}
\fbox{
\begin{picture}(95,95)
\Line(20,20)(80,20)
\Line(20,50)(80,50)
\Photon(50,20)(50,50){2}{6}
\Photon(35,50)(45,60){2}{2}
\Photon(45,60)(50,80){2}{3}
\Photon(45,60)(65,65){2}{3}
\end{picture}
}
\\\smallskip
\fbox{
\begin{picture}(95,95)
\Line(20,20)(80,20)
\Line(20,50)(80,50)
\Photon(50,20)(50,50){2}{6}
\Photon(65,50)(75,60){2}{2}
\Photon(75,60)(80,80){2}{3}
\Photon(75,60)(95,65){2}{3}
\end{picture}
}
\fbox{
\begin{picture}(95,95)
\Line(20,20)(80,20)
\Line(20,80)(80,80)
\Photon(50,20)(50,40){2}{4}
\Photon(50,40)(50,60){2}{4}
\Photon(50,60)(50,80){2}{4}
\Photon(50,40)(80,40){2}{4}
\Photon(50,60)(80,60){2}{4}
\end{picture}
}
\fbox{
\begin{picture}(95,95)
\Line(20,20)(80,20)
\Line(20,80)(80,80)
\Photon(50,20)(50,40){2}{4}
\DashLine(50,40)(50,60){2}
\Photon(50,60)(50,80){2}{4}
\Photon(50,40)(80,40){2}{4}
\Photon(50,60)(80,60){2}{4}
\end{picture}
}
\fbox{
\begin{picture}(95,95)
\Line(20,20)(80,20)
\Line(20,60)(80,60)
\Photon(50,20)(50,60){2}{6}
\Photon(30,60)(50,80){2}{4}
\Photon(40,60)(60,80){2}{4}
\end{picture}
}
\fbox{
\begin{picture}(95,95)
\Line(20,20)(80,20)
\Line(20,60)(80,60)
\Photon(50,20)(50,60){2}{6}
\Photon(60,60)(80,80){2}{4}
\Photon(70,60)(90,80){2}{4}
\end{picture}
}
\\\smallskip
\fbox{
\begin{picture}(95,95)
\Line(20,20)(80,20)
\Line(20,60)(80,60)
\Photon(50,20)(50,60){2}{6}
\Photon(35,60)(55,80){2}{4}
\Photon(65,60)(85,80){2}{4}
\end{picture}
}
\fbox{
\begin{picture}(95,95)
\Line(20,40)(80,40)
\Line(20,60)(80,60)
\Photon(50,40)(50,60){2}{4}
\Photon(35,40)(55,20){2}{4}
\Photon(65,60)(85,80){2}{4}
\end{picture}
}
\fbox{
\begin{picture}(95,95)
\Line(20,40)(80,40)
\Line(20,60)(80,60)
\Photon(50,40)(50,60){2}{4}
\Photon(35,40)(55,20){2}{4}
\Photon(35,60)(55,80){2}{4}
\end{picture}
}
\fbox{
\begin{picture}(95,95)
\Line(20,40)(80,40)
\Line(20,60)(80,60)
\Photon(50,40)(50,60){2}{4}
\Photon(65,40)(85,20){2}{4}
\Photon(65,60)(85,80){2}{4}
\end{picture}
}
\fbox{
\begin{picture}(95,95)
\Line(20,20)(80,20)
\Line(20,80)(80,80)
\Photon(50,20)(50,50){2}{6}
\Photon(50,50)(50,80){2}{6}
\Photon(50,50)(80,40){2}{4}
\Photon(50,50)(80,60){2}{4}
\Vertex(50,50){4}
\end{picture}
}
\end{center}
\caption{\footnotesize
  Topologies of the Feynman diagrams
  for the reaction $e^-e^- \to W^-Z^0\nu_e e^-$.
}
\label{ff4}
\end{figure}
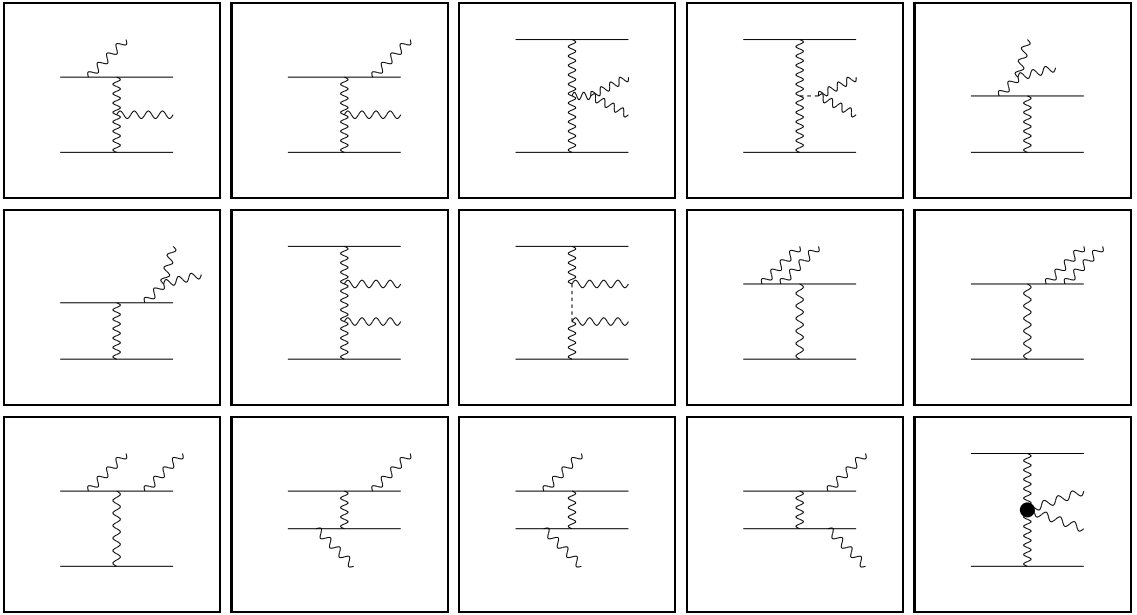

As we mentioned in the introduction,
the $L_1$ and $L_2$ quartic anomalous couplings
are expected to be larger
than the trilinear anomalies $L_{9L}$ and $L_{9R}$.
In a first approximation,
it is thus safe to ignore the latter
when it comes to study the former.
They can be tested to lowest order in the reactions
\bea
e^-e^- & \to & W^- W^- \nu_e \nu_e \\
&& W^- Z^0 \nu_e e^- \label{ee4}\\
&& Z^0 Z^0 e^- e^- \\
&& W^+ W^- e^- e^- ~.
\eea
It turns out
that is the most sensitive process to $L_1$ and $L_2$
is the second one (\ref{ee4}),
with both initial electron beams left polarized~\cite{ck}.
We therefore concentrate solely on this reaction from now on.
The 15 topologies of the 88 Feynman diagrams
participating to this reaction
in the unitary gauge
are shown in Fig.~\ref{ff4}.
It is the last diagram,
whose quartic vertex is emphasized,
which provides the signals we wish to test.

The expected \sm\ \xs s~\cite{ckr}
are displayed as functions of the \cm\ energy in Fig.~\ref{eny},
with the same cuts (\ref{cuts}) as in the analysis of the trilinear couplings,
except that in order to satisfy the requirement
that all events be fully reconstructible,
the efficiency drops to approximately 51\%.
Still,
substantial event rates are expected at high energy.

\begin{figure}[htb]
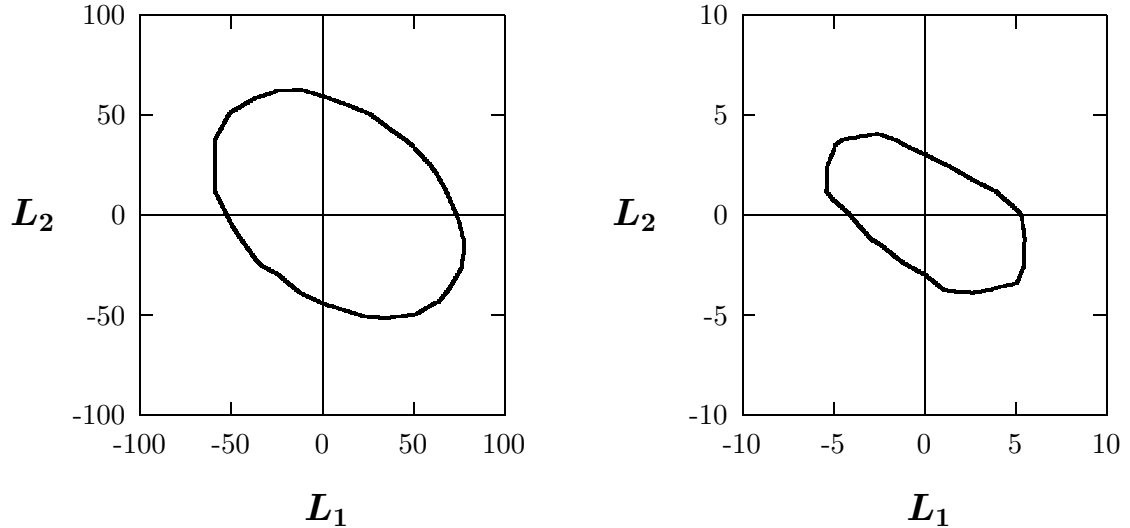

\vspace*{-8mm}
\boldmath
\centerline{\small
\input{q500.tex}
\input{q2000.tex}
}
\unboldmath
\caption{\footnotesize
  Contours of detectability
  at 95\%\ confidence
  for the quartic anomalous couplings
  in the $e^-_Le^-_L \to e^-\nu_eW^-Z^0$ reaction.
  The left plot is obtained
  for the \cm\ energy $\protect\sqrt{s}=500$ GeV
  and the luminosity ${\cal L}=20$ fb$^{-1}$,
  whereas the right plot is obtained with
  $\protect\sqrt{s}=2$ TeV and ${\cal L}=320$ fb$^{-1}$.
}
\label{qf}
\end{figure}

Here again,
we estimate the discovery potential of the reaction (\ref{ee4})
with the help of a least squares estimator.
As in the analysis of the trilinear couplings,
we assume a 2\%\ systematic error in this experiment too.
Because statistics are lower,
though,
we refrain from binning the data
and only use the total \xs.
Since modifications to the quartic couplings
destroy the very subtle and powerful unitarity cancellations
which in the \sm\ render the \xs\ well behaved,
this simple minded observable
may after all be rather efficient.
More work,
though,
should be performed in this direction
to confirm this working hypothesis.
Especially at higher energies,
where statistics become good,
the study of differential distributions
may significantly improve these results.

The results of the analysis
are plotted in Fig.~\ref{qf}
as the contours around the \sm\ expectation
beyond which the quartic anomalous couplings $L_1$ and $L_2$
can be excluded with 95\%\ confidence
in the absence of a signal.
Expectations are shown
for 500 and 2000 GeV
\cm\ energies.

\section{Conclusions}

We have analyzed the indirect effects
of a strongly interacting Higgs sector
in $e^-e^-$ scattering,
and find that $W^-$ production is sensitive
to the dominant anomalous trilinear and quartic gauge couplings.
The resolving power of these reactions
is comparable to the one expected from
$e^+e^-$, $e^-\gamma$ or $\gamma\gamma$ experiments
when a similar analysis is performed~\cite{cc,ck}.

The sensitivity to the anomalous couplings
increases substantially with the collider energy,
but a careful choice of sensitive observables remains essential
in order to obtain performing bounds.
Such optimizations,
including more elaborate treatments of the final states,
have been shown to dramatically improve the resolving power
of the $e^+e^-$ reactions~\cite{b}.
They are still to be performed for the $e^-e^-$ processes,
but there is no doubt that
there is a lot of room for improvement here too.

There were only four independent anomalous couplings
involved in this study,
because we restricted ourselves
to a heavy Higgs scenario.
If we relax this assumption
and settle for no less than any kind of new physics at the TeV scale,
a much larger number of anomalies
must be considered,
about twelve of lowest dimensions.
Having this in mind,
it becomes clear that a single experiment is not sufficient
to disentangle the complicated interdependences of all the parameters.
Therefore,
although none of the different
$e^+e^-$, $e^-e^-$, $e^-\gamma$ or $\gamma\gamma$
\lc\ operating modes is clearly performing better than the others
in probing anomalous couplings,
it will be important to gather as much information as possible
from all these experiments
in order to obtain the best resolution.

\section*{Acknowledgements}

It is a pleasure to thank Debajyoti Choudhury and Karol Ko\l odziej
for their collaboration on similar topics.

\end{document}